# CODED MODULATION for POWER LINE COMMUNICATIONS
## In AEÜ Journal, 2000, pp. 45-49, Jan 2000

A.J. Han Vinck

University of Essen
Ellernstr. 29, 45326, Essen
vinck@exp-math.uni-essen.de

**Abstract:** We discuss the application of coded modulation for power-line communications. We combine M-ary FSK with diversity and coding to make the transmission robust against permanent frequency disturbances and impulse noise. We give a particular example of the coding/modulation scheme that is in agreement with the existing CENELEC norms. The scheme can be considered as a form of coded Frequency Hopping and is thus extendable to any frequency range.

## INTRODUCTION

Power Line Communications (PLC) can be seen as one of the possible solutions to the *"last dirty mile"* problem for communication providers. However, there are several obstacles: 1) standards; 2) channel characteristics like attenuation, permanent frequency disturbances and impulsive noise; 3) network conditions. This paper focuses on the low frequency range below 150 kHz, where the CENELEC norms apply. According to the CENELEC norms, EN 50065.1, part 6.3.2, the maximum allowed peak voltage for narrow band transmitter output voltages at 9 kHz equals 5 V, exponentially decreasing to 1 V at 95 kHz; and for broad-band transmitters equals 5 V = 134 dB ($\mu$V). As a consequence, the transmitters are output voltage limited and bandwidth limited. Here, for narrow band signals the maximum signal power spectral density (PSD) within a bandwidth of 5 kHz is 20 dB larger than the signal PSD at the edges of the particular frequency band, i.e. a 20-dB bandwidth of less than 5 kHz in width. Broadband signals are defined as signals with a 20-dB bandwidth of more than 5 kHz in width.

The candidate modulation schemes with a constant envelope signal modulation such as binary-FSK and M-ary FSK are in agreement with to the CENELEC norms. A successful binary modulation scheme called Spread-FSK can be found in Schaub, [1]. This scheme is particularly suited for channels with non white interferers and/or channels with different attenuations at the two signaling frequencies. Another interesting multi carrier modulation scheme is Orthogonal Frequency Division Modulation ( OFDM, see [7] ). OFDM based systems are being developed for high speed transmission in frequency bands above 100 kHZ. Special algorithms are developed to reduce the maximum signal amplitude of the modulator output, see [8]. Both schemes can be combined with block or convolutional error control coding. Recent developments in modem design for PLC can be found in [9].

Channel characteristics regarding attenuation and noise have been reported in [2] and [3]. Television sets or computer terminals generate *narrow band* noise and thus, this type of noise is permanent over a long period of time and of great importance in power line communication systems. *Impulse noise* has been reported in [3]. From [3] it can be concluded that impulses have a duration of typically less than 100 $\mu$sec. However, more information about the statistical behavior of the impulse duration and inter-arrival times is needed. Measurements in networks indicate that the inter-arrival times are independent and .1 to 1 second apart. A modulation /coding scheme that incorporates frequency- and time diversity can be expected to be robust against both types of disturbances.

The main goal of the paper is to show that a combination of M-ary FSK modulation and coding can provide for a constant envelope modulation signal, frequency spreading to avoid bad parts of the frequency spectrum, and redundancy to facilitate correction of frequency disturbances and impulse noise simultaneously. M-ary FSK has the advantage of a constant envelope signal modulation and a demodulation in a coherent as well as a non-coherent way, leading to low complexity transceivers. In an M-ary FSK modulation scheme, symbols are modulated as one of the sinusoidal waves described by

$$s_i(t) = \sqrt{\frac{2E_s}{T_s}} \cos(2\pi f_i t); \quad 0 \leq t \leq T_s \qquad (1)$$

where $i = 1, 2, \cdots, M$ and $E_s$ is the signal energy per modulated symbol and

$$f_i = f_0 + \frac{i-1}{T_s}, \quad 1 \leq i \leq M.$$

The signals are orthogonal and for non-coherent reception the frequencies are spaced by $1/T_s$ Hz, being the transmission rate. To avoid abrupt switching from one frequency to another, the information bearing signal may modulate a single carrier whose frequency is changed continuously. The resulting frequency-modulated signal is phase continuous and is called continuous-phase FSK, (CPFSK). Details regarding this type of modulation and demodulation can be found in [6]. We restrict ourselves to the ideal M-FSK modulation as given in (1) and do not further consider spectral properties of the modulation scheme. According to [6] the theoretical measure of bandwidth efficiency, in bits/s/Hz, of M-FSK modulation is given by

$$\rho = \frac{\log_2 M}{M}.$$

For large M, M-ary FSK is spectrally inefficient. The *symbol* error probability for transmission over an Additive White Gaussian Noise (AWGN) channel with single sided noise power spectral density $N_0$, at high values of $E_s/N_0$, can be approximated as

$$P_s \approx \frac{1}{2} e^{-\frac{E_s}{2N_0}}, \qquad (2)$$

where $E_s = E_b \log_2 M$ and $E_b$ is the energy per information bit. For AWGN channels the probability of bit error can be made arbitrarily small by increasing M, provided that the signal-to-noise ratio (SNR) per bit $E_b/N_0$ is greater than the Shannon limit of $-1.6$ dB. The cost for increasing M is the bandwidth required to transmit the signals. For AWGN channels, the demodulation may be accomplished using 2M correlators, 2 per signal waveform. The optimum non-coherent demodulator detects M envelopes and outputs as estimate for the transmitted frequency the one that corresponds to the <u>largest</u> envelope. This makes the demodulation process simple and independent from the channel attenuation. However, as was shown in [1], this type of FSK demodulation is not optimum in case of a simplified model of a frequency selective channel as is the power line channel, where also signal distortions are neglected. In addition, the presence of narrow band noise may cause large envelopes and thus errors to occur at the demodulator output. Impulse noise has a broad band character and thus could lead to a multiple of large envelopes. To be able to handle these types of noise processes, we propose to modify the demodulator in such a way that the detected envelopes can be used in the decoding process of a special class of error control codes, called permutation codes. The combination of M-FSK with permutation coding gives rise to a constant modulator output envelope and includes frequency- and time diversity. Frequency- and time diversity can be expected to give robustness against narrow band noise and impulse noise. We first describe the properties of the combined modulation/coding scheme. Then we give the modification of the demodulator and the influence of the channel noise on the demodulator output.

**COMBINED MODULATION and CODING**
Consider an M-ary FSK signal set that consists of an orthogonal set of M frequency-shifted signals. We may assume non-coherent demodulation and thus the bandwidth required to transmit M frequencies is approximately

$$B = \frac{M}{T_s},$$

where $T_s$ is the signal duration time. Note we transmit only a single frequency at a time and thus the transmitted signal has a constant envelope output. In the following we use the integers $1, 2, \cdots, M$ to represent the M frequencies, i.e. the integer i represents $f_i$. A message is encoded as a code word of length M with the integers 1,2, ⋯, M as symbols. The symbols of a code word are transmitted in time as the corresponding frequencies. Let |C| denote the cardinality of the code.

*Definition.* A permutation code C consists of |C| code words of length M, where every code word contains M <u>different</u> symbols.

Example: M = 4, |C| = 4.

| message | transmit |
|---------|----------|
| 1 | ( 1, 2, 3, 4 ) |
| 2 | ( 2, 1, 4, 3 ) |
| 3 | ( 3, 4, 1, 2 ) |
| 4 | ( 4, 3, 2, 1 ) |

As an example, message 3 is transmitted in time as the series of frequencies ( $f_3$, $f_4$, $f_1$, $f_2$ ). Note that the code C has 4 words with the property that two words always differ in 4 positions.

**TABLE 1**

Code book for M = 4, $d_{min}$ = 3

| Message | Code word |
|---------|-----------|
| 1 | 1, 2, 3, 4 |
| 2 | 1, 3, 4, 2 |
| 3 | 2, 1, 4, 3 |
| 4 | 2, 4, 3, 1 |
| 5 | 3, 1, 2, 4 |
| 6 | 3, 4, 1, 2 |
| 7 | 4, 2, 1, 3 |
| 8 | 4, 3, 2, 1 |
| 9 | 1, 4, 2, 3 |
| 10 | 2, 3, 1, 4 |
| 11 | 3, 2, 4, 1 |
| 12 | 4, 1, 3, 2 |

The code as given in table 1 has 12 words, each with M=4 different numbers and a minimum difference between any two words or *minimum Hamming distance* $d_{min}$ of 3. For M = 3 we have the code books as given in table 2.

**TABLE 2**

Two code books for M = 3

| $d_{min}$ = 2 | $d_{min}$ = 3 |
|---------------|---------------|
| 1 2 3 | 1 2 3 |
| 1 3 2 | |
| 2 1 3 | 2 3 1 |
| 2 3 1 | |
| 3 1 2 | 3 1 2 |
| 3 2 1 | |

An interesting problem is the design of codes and the effect of coding on the transmission efficiency. In table 3 we give the code construction results for M < 6.

TABLE 3

Code book sizes for M = 2, 3, 4, 5

| M | $d_{min}=$ | | | |
|---|---|---|---|---|
| | 2 | 3 | 4 | 5 |
| 2 | 2 | | | |
| 3 | 6 | 3 | | |
| 4 | 24 | 12 | 4 | |
| 5 | 120 | 60 | 20 | 5 |

If we have a information transmission rate of b bits per second, we obtain a signal duration time

$$T_s = \frac{1}{b} \cdot \frac{\log_2 |C|}{M} . \tag{3}$$

The bandwidth required is thus approximately

$$B = M \cdot \frac{b \cdot M}{\log_2 |C|} . \tag{4}$$

The bandwidth efficiency of this coded M-ary FSK scheme is defined by

$$\rho = \frac{b}{B} = \frac{\log_2 |C|}{M^2} . \tag{5}$$

To maximize the efficiency, we have to find the largest |C| for a given M and $d_{min}$. It is easy to see that for a code with code words of length M each having M different numbers $d_{min}$ is always $\geq 2$. The cardinality |C| of this code is M!. Hence, the bandwidth efficiency can be defined as

$$\rho = \frac{\log_2 M!}{M^2} \approx \frac{\log_2 M}{M} , \tag{6}$$

for large M. This is the same efficiency as uncoded M-ary FSK. The next theorem gives an upper bound on the number of code words in a permutation code.

*Theorem.* For a permutation code of length M with M different code symbols in every code word and minimum Hamming distance $d_{min}$, the cardinality is upper bounded by

$$|C| \leq \frac{M!}{(d_{min}-1)!} . \tag{7}$$

Proof. Choose ( M - $d_{min}$ ) different code symbols and put them at ( M - $d_{min}$ ) different positions. Then, with the remaining $d_{min}$ symbols at the remaining $d_{min}$ positions, we can construct a maximum number of $d_{min}$ code words at a distance $d_{min}$ apart. For every set of positions we can distribute the chosen ( M - $d_{min}$ ) different code symbols in ( M-$d_{min}$ )! different ways. We can choose $\binom{M}{M-d_{min}}$ different sets of positions. Hence, the overall product gives the maximum number of code words with a minimum Hamming distance $d_{min}$,

$$|C| \leq d_{min} \binom{M}{M-d_{min}} ( M - d_{min} )! = \frac{M!}{(d_{min}-1)!} .$$

It can be shown mathematically, that for M = 6 and $d_{min}$ = 5 the upper bound (7) cannot be met with equality. For $d_{min}$ = 2, we always have equality in (7) for any M. Blake, [5], uses the concept of sharply k-transitive groups to define permutation codes with distance M – k + 1. The structure of all sharply k-transitive groups is known for k ≥ 2. In [5] it is also shown that the group of even permutations on M letters is a permutation code with |C| = M!/2 code words and minimum Hamming distance 3. To find good codes in general appears to be quite difficult. The codes described in [5] are "simple" examples. If we assume that codes exist that meet (7) with equality, then from (5) it follows that the bandwidth efficiency defined for a code with minimum Hamming distance $d_{min}$ and increasing length M approaches

$$\rho \approx \frac{M - d_{min} + 1}{M} \cdot \frac{\log_2 M}{M}. \qquad (8)$$

In the next section we discuss the modification of the demodulator and the use of the modified demodulator output in the decoding of permutation codes.

**DEMODULATION and DECODING**
Following [1], the conventional non-coherent demodulator is preceded by an automatic gain control unit (AGC). The demodulation may be accomplished using 2M correlators, 2 per signal waveform. The sub-optimum non-coherent demodulator computes M envelopes and outputs as estimate for the transmitted frequency the one that corresponds to the largest envelope. An optimum decision rule can be derived using the knowledge of the SNR per sub-channel for a particular frequency. In practical schemes the SNR per sub-channel can be obtained with the help of a well defined preamble or by using the output of the correlators, see also [1].

As indicated before, spectral noise disturbs the demodulation process of M-ary FSK. To overcome this problem, we give a possible structure for a sub-optimum demodulator using a threshold after every envelope detector. We modify the demodulator for our modulation/coding scheme as follows. Suppose that after every envelope detector we introduce a threshold $T_i$, i = 1,2, ⋯ , M. For practical reasons the value of $T_i$ is set equal to $\sqrt{E_{s,i}}/2$, where $\sqrt{E_{s,i}}$ is the envelope detector output for channel i if no noise is present. For noisy channels, we have to add a positive constant, which incorporates the variance of the noise, to the threshold. The symbols (frequencies) corresponding to the envelopes that are larger than the respective thresholds are given as demodulator outputs. Thus, the inputs to the decoder are then multi-valued. The decoder acts according to the following decoding rule.

*Decoding rule:* Output the message corresponding to the code word that has the maximum number of agreements with the symbols (frequencies) at the demodulator output.

For the example with M = 4 and $d_{min}$= 4, a permanent disturbance (narrow band noise) present at the sub-channel for frequency $f_4$ and transmission of code word { 3, 4, 1, 2 } could lead to a demodulator output { (3,4), (4), (1,4), (2,4) }. The decoder compares the demodulator output with all possible transmitted code words. It outputs the code word for which the maximum number of agreements with the symbols at the demodulator output occurs. For the example, all symbols corresponding to code word 3 are present and thus correct decoding follows. Since code words are different in at least $d_{min}$ positions, $d_{min}$ – 1 errors of this type still allow correct decoding. The example code has $d_{min}$ = 4 and hence, we can tolerate the presence of 3 permanent disturbances present in the demodulator output.

Impulse noise has a broad frequency band character and has been reported in [3]. From this it can be concluded that impulses have a duration of typically less than 100 μsec. However, more information about the statistical behavior of the impulse duration and inter-arrival times is needed. Measurements in networks indicate that the inter-arrival times are independent and .1 to 1 second apart. For a signaling scheme using a signaling rate of 10kHz, we have a symbol duration of 100μsec, which is in the range of the impulse duration. So, impulse noise, affecting at least two adjacent symbols cannot be excluded. Due to the broad frequency band character, impulse noise may cause the demodulator to output the presence of all frequencies. This type of noise can be seen as erasures. Hence, two affected adjacent transmissions may reduce the minimum distance of the code by 2. At a signaling rate of 10 kHz, a code with $d_{min}$ = 3 is thus capable of correcting 2 permanent disturbances or to give a correct output in the presence of an impulse. For higher signaling rates, more symbols may be affected and thus a code with a larger minimum distance is needed. Suppose that we transmit the example code word {3, 4, 1, 2}. If an impulse noise causes all envelopes to be present at three symbol transmissions, then we may have as a demodulator output { (1,2,3,4), (1,2,3,4), (1,2,3,4), (2) }. Comparing this output with the possible transmitted code words gives a difference (distance) of zero to the correct code word and one to all other code words. Thus, even if three of these multi-valued outputs occur, we are still able to find back the correct code word since there is always a remaining symbol that gives a difference of one to the incorrect code words.

Background noise degrades performance by introducing unwanted (called insertions) demodulator outputs or by causing the absence (called deletion) of a transmitted frequency in the demodulator output. Note that for this type of *"threshold"* demodulation, the decoding is still correct for $d_{min} - 1$ errors of the insertion/deletion type.

- The absence of a frequency in the demodulator output always reduces the number of agreements between a transmitted code word and the received code word by one. The same is true for the other code words having the same symbol at the same position. If the symbols are different the number of agreements does not change.
- The appearance of every unwanted output symbol may increase the number of agreements between a wrong code word and the received code word by one. It does not decrease the number of agreements between a transmitted code word and the received code word.

In conclusion, we can say that the introduced thresholds in the modified demodulator in combination with the permutation code allow the correction of $d_{min} - 1$ incorrect demodulator outputs caused by narrow band-, impulse- or background noise. It can be shown [1] that the probability of an insertion or deletion error due to background noise is of the same type as for simple Amplitude Shift Keying (ASK). In fact, the bit error probability for the indicated threshold setting is approximated as

$$P_{e,ASK} \approx 0.5 \exp(-SNR_i / 4), \tag{9}$$

where $SNR_i$ is the SNR for channel i defined as

$$SNR_i = \frac{E_{s,i}}{N_i},$$

and $N_i$ is the single sided noise power spectral density for sub-channel i. From [2] it can be concluded that the received signal power $S_{re}$ for a bad transmission channel is given by

$$S_{re} = S_{in} \cdot 10^{-0.01 \cdot L}, \tag{10}$$

where $S_{in}$ is the transmitter output power and L the distance between transmitter and receiver in meters. According to the CENELEC norms, we take $S_{in} = 25$ Watt, and we may use a bandwidth up to 95 kHz. The worst case single sided noise power spectral density is given by:

$$N_{re, worst}(f) = 10^{-8 - 4 \cdot 10^{-5} \cdot f}. \tag{11}$$

Assuming that the received signal energy is the same for all sub-channels, the received signal energy and received power are related as $E_s = S_{re} T_s$. We give an example of an M-ary FSK scheme for M = 4 with an information rate of b = 4.8 kb/s. This scheme has the following properties:

- $B = \frac{4.8 \cdot 16}{\log_2 |C|}$ kHz . From here on, B is given in kHz.
- we divide the frequency band up to 95 kHz into 4 sub-bands each of bandwidth B/4
- we let the input power spectral density constant

The total power used at a distance of 500 meters (10) is $25 \cdot 10^{-5}$ W. The noise power spectral density (11) at (95-B) kHz is $N_{re} = 10^{-8 - 0.04(95 - B)}$ W/Hz. Using the 4-tone bandwidth of B kHz, we then obtain a signal to noise ratio lower bounded by

$$SNR_i > \frac{25 \cdot 10^{-5}}{(B/4) \cdot 10^3 \cdot 10^{-8 - 0.04(95 - B)}}.$$

Table 4 gives an overview of the obtained results.

**TABLE 4**

SNR in dB as a function of $d_{min}$ for M = 4

| $d_{min}$ | B | SNR |
|---|---|---|
| 2 | 16 kHz | 40 dB |
| 3 | 21 kHz | 37 dB |
| 4 | 38 kHz | 27 dB |

The size of the code books depends on the minimum Hamming distance between the code words. Since we fixed the information rate, the signal-to-noise ratio in Table 4 depends on the bandwidth B. For $d_{min}$ = 4, the signaling scheme can tolerate up to 3 frequency disturbances and the information rate equals 4.8 kb/s. Using (9), one sees that the error probability due to back ground noise can be neglected and is of no importance at all.

An additional option that can be implemented is to consider the transmitted code words as symbols of a longer code. We may for instance use a Reed Solomon (RS) code, see [4]. Symbol (code word) errors are then corrected by the RS code. Of course, the code symbols of the RS code need to be code words of our permutation code. The parameters of such a concatenated coding scheme depends on the particular application.

**CONCLUSIONS**

We describe a new modulation/coding scheme that is capable of handling frequency disturbances like narrow band noise or impulse noise. For this we use the concept of M-FSK modulation combined with permutation codes. We modify the "simple" non-coherent demodulator in such a way that the demodulator output becomes multi-valued. The demodulator output is used in the decoding of the permutation code. It appears that background noise is of no importance up to a distance of 500 meters. The described modulation/coding scheme leads to an overall robust system.